\documentclass[english,pccp,preprint,superscriptaddress]{revtex4-1} 
\usepackage{epsfig}
\usepackage{amsmath} 
\usepackage{graphicx} 
\usepackage{longtable} 
\usepackage{hyperref} 
\usepackage{amssymb} 
\usepackage{dcolumn} 
\usepackage{mhchem} 
\usepackage{cleveref} 
\usepackage{subfigure} 
\usepackage{array} 
\usepackage{multirow} 
\usepackage{tabularx} 
\usepackage{cleveref} 
\begin{document} 

\title{Effect of Li Termination on the Electronic and Hydrogen Storage Properties of Linear Carbon Chains: A TAO-DFT Study} 

\author{Sonai Seenithurai} 
\affiliation{Department of Physics, National Taiwan University, Taipei 10617, Taiwan} 

\author{Jeng-Da Chai} 
\email[Author to whom correspondence should be addressed. Electronic mail: ]{jdchai@phys.ntu.edu.tw} 
\affiliation{Department of Physics, National Taiwan University, Taipei 10617, Taiwan} 
\affiliation{Center for Theoretical Sciences and Center for Quantum Science and Engineering, National Taiwan University, Taipei 10617, Taiwan} 

\date{\today} 

\begin{abstract} 

Accurate prediction of the electronic and hydrogen storage properties of linear carbon chains (C$_{n}$) and Li-terminated linear carbon chains (Li$_{2}$C$_{n}$), with $n$ carbon 
atoms ($n$ = 5--10), has been very challenging for traditional electronic structure methods, due to the presence of strong static correlation effects. To meet the challenge, we study 
these properties using our newly developed thermally-assisted-occupation density functional theory (TAO-DFT), a very efficient electronic structure method for the study of large 
systems with strong static correlation effects. Owing to the alteration of the reactivity of C$_{n}$ and Li$_{2}$C$_{n}$ with $n$, odd-even oscillations in their electronic properties are 
found. In contrast to C$_{n}$, the binding energies of H$_{2}$ molecules on Li$_{2}$C$_{n}$ are in (or close to) the ideal binding energy range (about 20 to 40 kJ/mol per H$_{2}$). 
In addition, the H$_{2}$ gravimetric storage capacities of Li$_{2}$C$_{n}$ are in the range of 10.7 to 17.9 wt\%, satisfying the United States Department of Energy (USDOE) 
ultimate target of 7.5 wt\%. On the basis of our results, Li$_{2}$C$_{n}$ can be high-capacity hydrogen storage materials for reversible hydrogen uptake and release 
at near-ambient conditions. 

\end{abstract} 

\maketitle

\section*{Introduction} 

Hydrogen (H$_{2}$), as a pure energy carrier, has many attributes. Being light weight, it carries 142 MJ/kg of energy, which is approximately three times the energy content of gasoline, 
in terms of mass. Also, it is highly abundant on the earth in the form of water. More importantly, when hydrogen is burned with oxygen, it releases water vapor as the only effluent. 
Despite these advantages, there remain several problems to be clarified for the use of hydrogen. For example, hydrogen is highly flammable, and hence, if it comes in contact with the 
environment, it will burst. Another problem is related to its low energy content in terms of volume: it has only 0.0180 MJ/L, which is very low relative to gasoline (34.8 MJ/L). Moreover, 
over the past few years, the storage of hydrogen for onboard applications has been an active arena, which also requires a lightweight storage medium. Because of these reasons, 
storing a large amount of hydrogen reversibly in a small and lightweight container safely has been the biggest challenge in realizing a hydrogen-based 
economy \cite{Schlapbach2001,Jena2011,Park2012,Dalebrook2013,usdoe}. 

Over the years, the United States Department of Energy (USDOE) has monitored the research progress in the development of hydrogen storage materials for consumer vehicles. 
In 2015, the USDOE set the ultimate target of 7.5 wt\% for the gravimetric storage capacities of onboard hydrogen storage materials for light-duty vehicles \cite{usdoe}. As of now, 
there have been several methods for the storage of hydrogen \cite{Schlapbach2001,Jena2011,Park2012,Dalebrook2013}. The conventional methods for storing hydrogen are the high 
pressure method and the cryogenic method. In the high pressure method, one adopts carbon fiber reinforced tanks, which can withstand very high pressures (e.g., 350 to 700 bar), to 
store a large amount of completely recoverable hydrogen. In the cryogenic method, hydrogen is stored at very low temperatures (e.g., 20 K), typically requiring an expensive liquid 
helium refrigeration system. Both of these methods are not suitable for onboard automobile applications, because of the associated risk, high cost, and heavy weight. The storage of 
hydrogen in a metal hydride seems to be a convincing solution, but the irreversibility, slow kinetics, and high desorption temperature associated with this method are the problems yet 
to be overcome. Another promising solution is the storage of hydrogen in high surface area materials (e.g., graphene, carbon nanotubes, and metal-organic frameworks) through the 
adsorption-based methods. As high surface area materials could adsorb large amounts of hydrogen, the corresponding H$_{2}$ gravimetric storage capacities could be rather high. 
Nevertheless, these materials bind H$_{2}$ molecules very weakly (i.e., mainly governed by van der Waals (vdW) interactions), and hence, they perform properly only at low 
temperatures. 

For reversible hydrogen adsorption and desorption at ambient conditions (298 K and 1 bar), the ideal binding energies of H$_{2}$ molecules on hydrogen storage materials should 
be in the range of about 20 to 40 kJ/mol per H$_{2}$ \cite{Bhatia2006,Lochan2006,Sumida2013}. Consequently, various novel methods are being explored to increase the binding 
energies of H$_{2}$ molecules on high surface area materials to the aforementioned ideal range for ambient storage applications. To increase the H$_{2}$ adsorption binding energy, 
the surface of the adsorbent is generally modified with substitution doping, adatom adsorption, functionalization, etc. \cite{Jena2011}. Among them, Li adsorption is especially attractive, 
because of its light weight with which a high gravimetric storage capacity could be easily achieved. Note also that Li-adsorbed carbon materials have been shown to possess relatively 
high gravimetric storage capacities with enhanced H$_{2}$ adsorption binding energies \cite{Chen1999,Deng2004,Deng2010,SeenithuraiLi,Qiu2014,TAOH2S1}, through 
a charge-transfer induced polarization mechanism \cite{Niu1992,Niu1995,Froudakis2001,Jena2011}. 

\begin{figure} 
\includegraphics[scale=0.5]{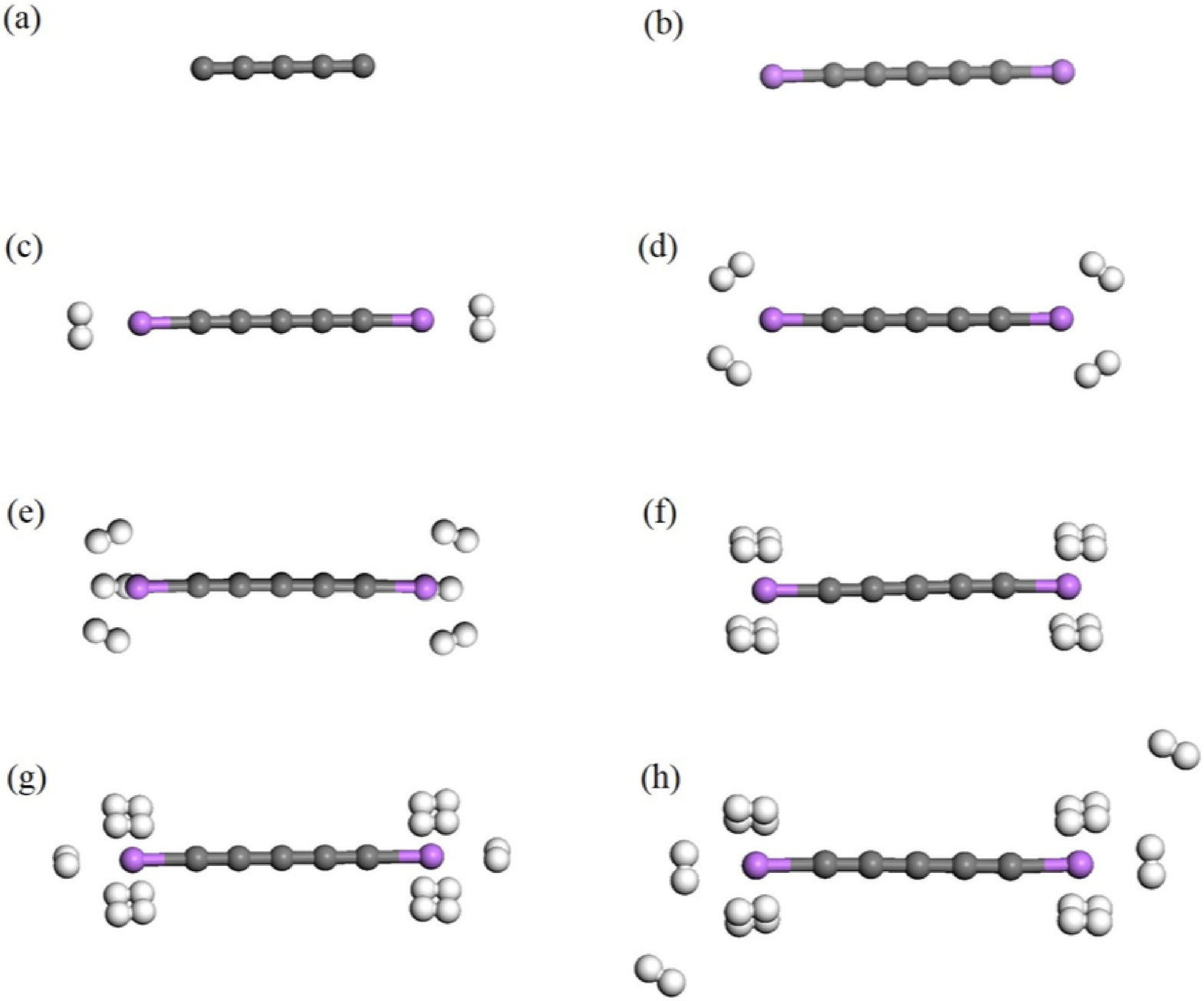} 
\caption{\label{fig:Figure_Geometry} 
Structures of (a) C$_{5}$, (b) Li$_{2}$C$_{5}$, (c) Li$_{2}$C$_{5}$ with one H$_{2}$ molecule adsorbed on each Li atom, 
(d) Li$_{2}$C$_{5}$ with two H$_{2}$ molecules adsorbed on each Li atom, (e) Li$_{2}$C$_{5}$ with three H$_{2}$ molecules adsorbed on each Li atom, 
(f) Li$_{2}$C$_{5}$ with four H$_{2}$ molecules adsorbed on each Li atom, (g) Li$_{2}$C$_{5}$ with five H$_{2}$ molecules adsorbed on each Li atom, and 
(h) Li$_{2}$C$_{5}$ with six H$_{2}$ molecules adsorbed on each Li atom. Here, grey, white, and purple balls represent C, H, and Li atoms, respectively.} 
\end{figure}

Among carbon materials, linear carbon chains (C$_{n}$), consisting of $n$ carbon atoms bonded with sp$^{1}$ hybridization (see \Cref{fig:Figure_Geometry}(a)), have recently 
attracted much attention owing to their unique electronic 
properties \cite{Banhart2015,Cesari2016,Jin2009,Chuvilin2009,Kano2014,VanZee1988,Pan2003,Fan1989,Heimann1999,Horny2002,Belau2007,Lang1998,Pan2003,Souza2008,
Li2009,Artyukhov2014,Banhart2015,Cesari2016}. Note that C$_{n}$ may be considered for hydrogen storage applications due to their one-dimensional (1D) structures and the 
feasibility of synthesis of C$_{n}$ and their derivatives \cite{Banhart2015,Cesari2016,Jin2009,Chuvilin2009,Kano2014,VanZee1988,Pan2003}. Recently, Pt-terminated linear carbon 
chains have been synthesized \cite{Kano2014}. As mentioned above, due to a charge-transfer induced polarization mechanism \cite{Niu1992,Niu1995,Froudakis2001,Jena2011}, 
Li-terminated linear carbon chains (Li$_{2}$C$_{n}$) can be good candidates for hydrogen storage materials at near-ambient conditions (see \Cref{fig:Figure_Geometry}(b--h)). 
Because of the light elements (i.e., C and Li atoms) in Li$_{2}$C$_{n}$, high gravimetric storage capacities could be easily achieved. However, to the best of our knowledge, there has 
been no comprehensive study on the electronic and hydrogen storage properties of Li$_{2}$C$_{n}$ in the literature, possibly due to the presence of strong static correlation effects 
in Li$_{2}$C$_{n}$ (commonly occurring in 1D structures due to quantum confinement effects \cite{Brus2014}). Theoretically, the popular Kohn-Sham density functional theory 
(KS-DFT) \cite{Kohn1965} with conventional semilocal \cite{PBE}, hybrid \cite{hybrid,wM05-D,LC-D3,SLC-D3}, and double-hybrid \cite{B2PLYP,wB97X-2,PBE0-2,SCAN0-2} 
exchange-correlation (XC) density functionals can provide unreliable results for systems with strong static correlation effects \cite{Cohen2012}. For accurate prediction of the properties 
of these systems, high-level {\it ab initio} multi-reference methods are typically needed \cite{multi-reference}. Nonetheless, accurate multi-reference calculations are prohibitively 
expensive for large systems (especially for geometry optimization). 

To circumvent the formidable computational expense of high-level {\it ab initio} multi-reference methods, we have newly developed thermally-assisted-occupation density functional theory 
(TAO-DFT) \cite{ChaiTAO2012,ChaiTAO2014,ChaiTAO2017} for the study of large ground-state systems (e.g., containing up to a few thousand electrons) with strong static correlation 
effects. In contrast to KS-DFT, TAO-DFT is a density functional theory with fractional orbital occupations, wherein strong static correlation is explicitly described by the entropy contribution 
(see Eq.\ (26) of Ref.\ \cite{ChaiTAO2012}), a function of the fictitious temperature and orbital occupation numbers. Note that the entropy contribution is completely missing in KS-DFT. 
Interestingly, TAO-DFT is as efficient as KS-DFT for single-point energy and analytical nuclear gradient calculations, and is reduced to KS-DFT in the absence of strong static correlation 
effects. Therefore, TAO-DFT can treat both single- and multi-reference systems in a more balanced way than KS-DFT. Besides, existing XC density functionals in KS-DFT may also be 
adopted in TAO-DFT. Due to its computational efficiency and reasonable accuracy for large systems with strong static correlation, TAO-DFT has been successfully applied to the study of 
several strongly correlated electron systems at the nanoscale \cite{Wu2015,NK,TAOH2S1,cycl}, which are typically regarded as ``challenging systems" for traditional electronic structure 
methods (e.g., KS-DFT with conventional XC density functionals and single-reference {\it ab initio} methods) \cite{Cohen2012}. Accordingly, TAO-DFT can be an ideal theoretical method 
for studying the electronic properties of Li$_{2}$C$_{n}$. Besides, the orbital occupation numbers in TAO-DFT can be useful for examining the possible radical character of 
Li$_{2}$C$_{n}$. For the hydrogen storage properties, as the interaction between H$_{2}$ and Li$_{2}$C$_{n}$ may involve dispersion (vdW) interactions, electrostatic interactions, 
and orbital interactions \cite{Lochan2006,Park2012,Tsivion2014}, the inclusion of dispersion corrections \cite{BLYP-D,Grimme2016} in TAO-DFT is important for properly describing 
noncovalent interactions. Therefore, in this work, we adopt TAO-DFT with dispersion corrections \cite{ChaiTAO2014} to study the electronic and hydrogen storage properties of 
Li$_{2}$C$_{n}$ with various chain lengths ($n$ = 5--10). In addition, the electronic properties of Li$_{2}$C$_{n}$ are also compared with those of C$_{n}$ to examine the role of 
Li termination.

\section*{Computational Details} 

All calculations are performed with a development version of \textsf{Q-Chem 4.4} \cite{Shao2015}, using the 6-31G(d) basis set with the fine grid EML(75,302), consisting of 
75 Euler-Maclaurin radial grid points and 302 Lebedev angular grid points. Results are computed using TAO-BLYP-D \cite{ChaiTAO2014} (i.e., TAO-DFT with the dispersion-corrected 
BLYP-D XC density functional \cite{BLYP-D} and the LDA $\theta$-dependent density functional $E_{\theta}^{\text{LDA}}$ (see Eq.\ (41) of Ref.\ \cite{ChaiTAO2012})) with the fictitious 
temperature $\theta$ = 7 mhartree (as defined in Ref.\ \cite{ChaiTAO2012}).

\section*{Results and Discussion} 

\subsection*{Electronic Properties} 

\begin{figure} 
\includegraphics[scale=0.66]{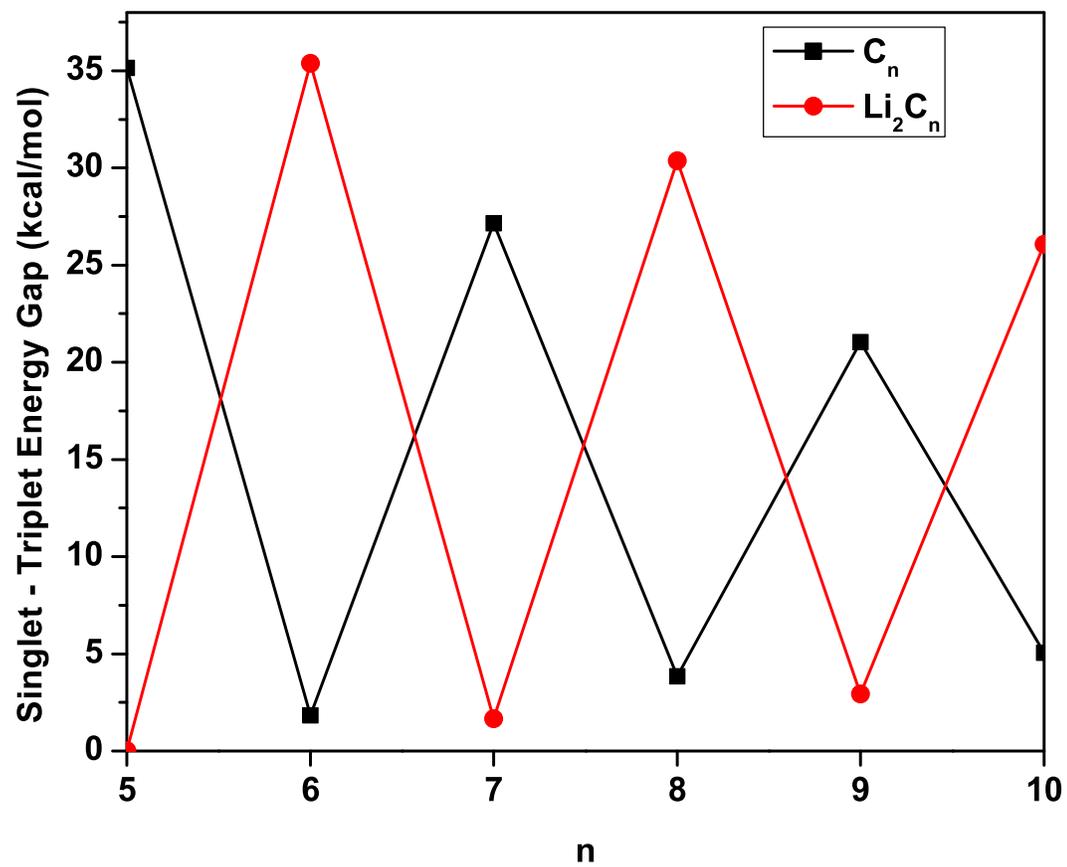} 
\caption{\label{fig:Figure_ST_Gap} 
Singlet-triplet energy (ST) gap of C$_{n}$/Li$_{2}$C$_{n}$ as a function of the chain length, calculated using TAO-BLYP-D.} 
\end{figure}

To obtain the ground state of C$_{n}$/Li$_{2}$C$_{n}$ ($n$ = 5--10), spin-unrestricted TAO-BLYP-D calculations are performed for the lowest singlet and triplet energies of 
C$_{n}$/Li$_{2}$C$_{n}$ on the respective geometries that were fully optimized at the same level of theory. The singlet-triplet energy (ST) gap of C$_{n}$/Li$_{2}$C$_{n}$ is 
calculated as $(E_{\text{T}} - E_{\text{S}})$, the energy difference between the lowest triplet (T) and singlet (S) states of C$_{n}$/Li$_{2}$C$_{n}$. 
As shown in \Cref{fig:Figure_ST_Gap}, the ground states of C$_{n}$ and Li$_{2}$C$_{n}$ are singlets for all the chain lengths investigated. 

Because of the symmetry constraint, the spin-restricted and spin-unrestricted energies for the lowest singlet state of C$_{n}$/Li$_{2}$C$_{n}$ should be the same for the exact 
theory \cite{Rivero2013,ChaiTAO2012,ChaiTAO2014,ChaiTAO2017}. To assess the possible symmetry-breaking effects, we also perform spin-restricted TAO-BLYP-D calculations 
for the lowest singlet energies on the corresponding optimized geometries. The spin-restricted and spin-unrestricted TAO-BLYP-D energies for the lowest singlet state 
of C$_{n}$/Li$_{2}$C$_{n}$ are found to be essentially the same (within the numerical accuracy of our calculations), implying that essentially no unphysical symmetry-breaking 
effects occur in our spin-unrestricted TAO-BLYP-D calculations. 

\begin{figure} 
\includegraphics[scale=0.66]{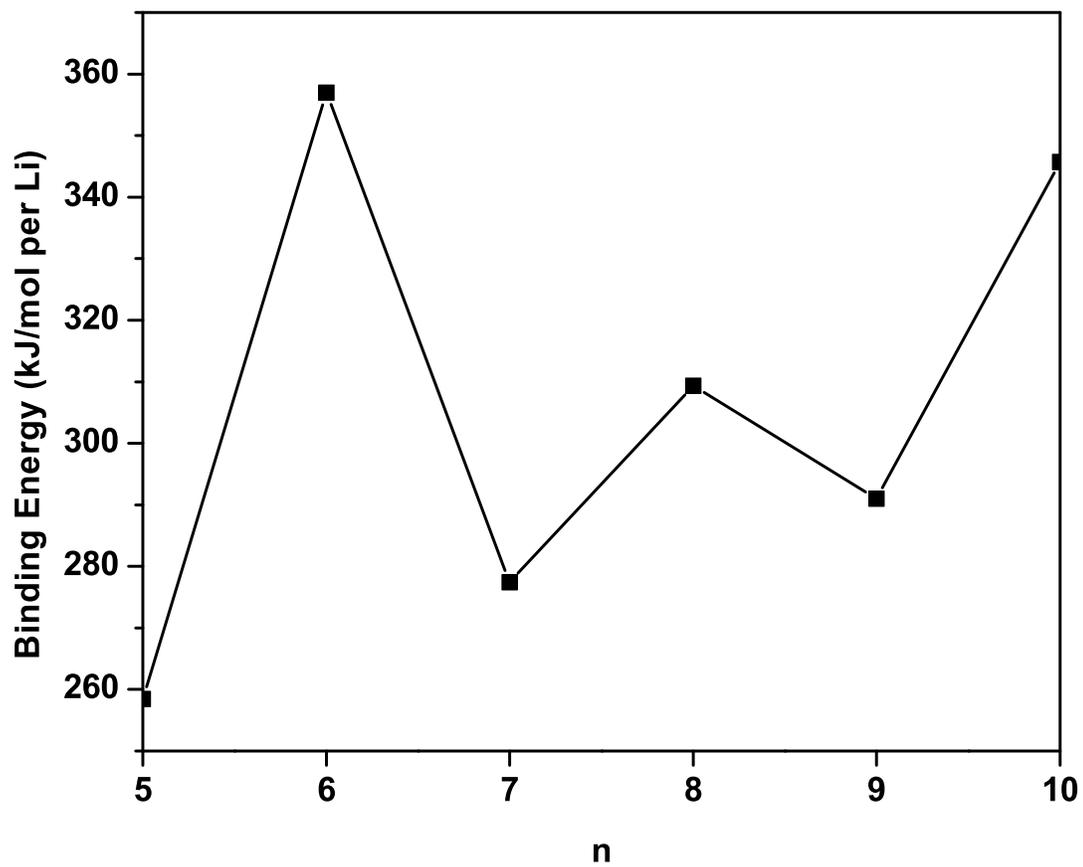} 
\caption{\label{fig:Figure_BE_of_Li} 
Li binding energy on C$_{n}$ as a function of the chain length, calculated using TAO-BLYP-D.} 
\end{figure}

To assess the energetic stability of terminating Li atoms, the Li binding energy, $E_{b}(\text{Li})$, on C$_{n}$ is computed using 
\begin{equation}\label{eq:EBLi} 
E_{b}(\text{Li}) = (E_{\text{C}_{n}} + 2 E_{\text{Li}} - E_{\text{Li}_{2}\text{C}_{n}}) / 2, 
\end{equation} 
where $E_{\text{C}_{n}}$ is the total energy of C$_{n}$, $E_{\text{Li}}$ is the total energy of Li, and $E_{\text{Li}_{2}\text{C}_{n}}$ is the total energy of Li$_{2}$C$_{n}$. 
$E_{b}(\text{Li})$ is subsequently corrected for the basis set superposition error (BSSE) using the counterpoise correction \cite{Boys1970}, 
where the C$_{n}$ is considered as one fragment, and the 2 Li atoms are considered as the other fragment. 
As shown in \Cref{fig:Figure_BE_of_Li}, C$_{n}$ can strongly bind the Li atoms with the binding energy range of 258 to 357 kJ/mol per Li. 

\begin{figure} 
\includegraphics[scale=0.66]{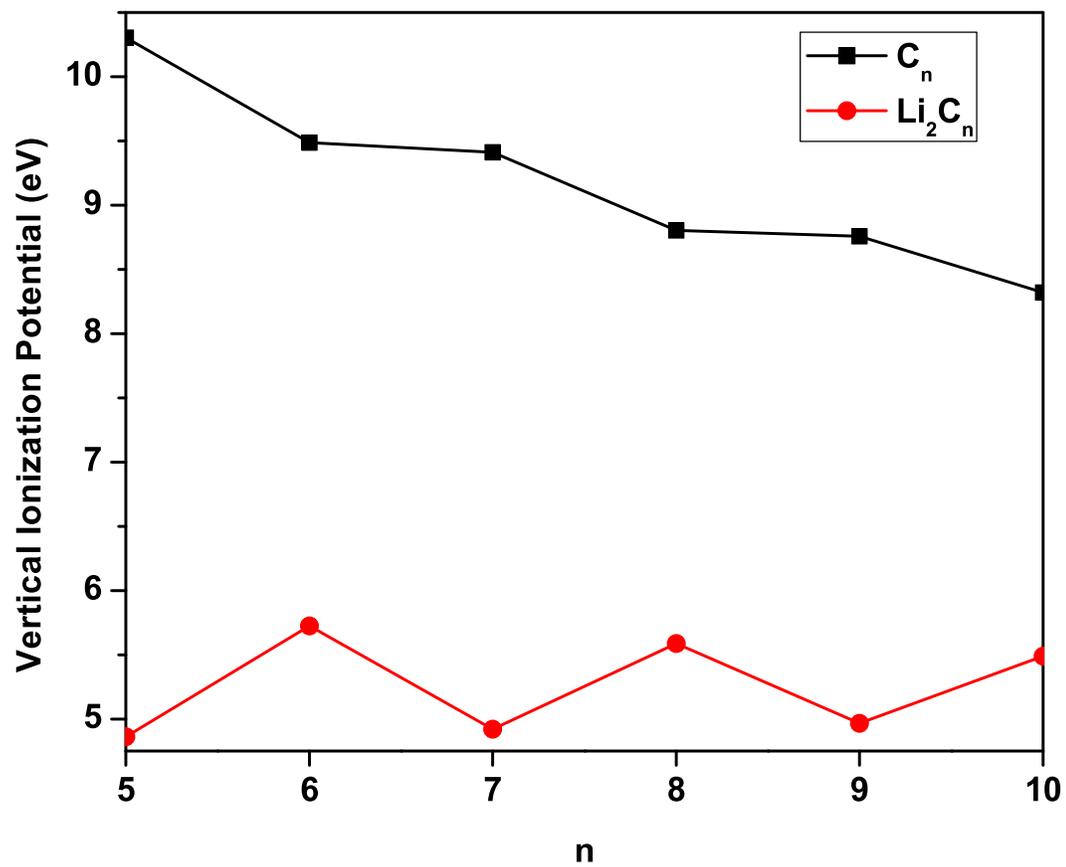} 
\caption{\label{fig:Figure_IPv} 
Vertical ionization potential for the lowest singlet state of 
C$_{n}$/Li$_{2}$C$_{n}$ as a function of the chain length, calculated using TAO-BLYP-D.} 
\end{figure}

\begin{figure} 
\includegraphics[scale=0.66]{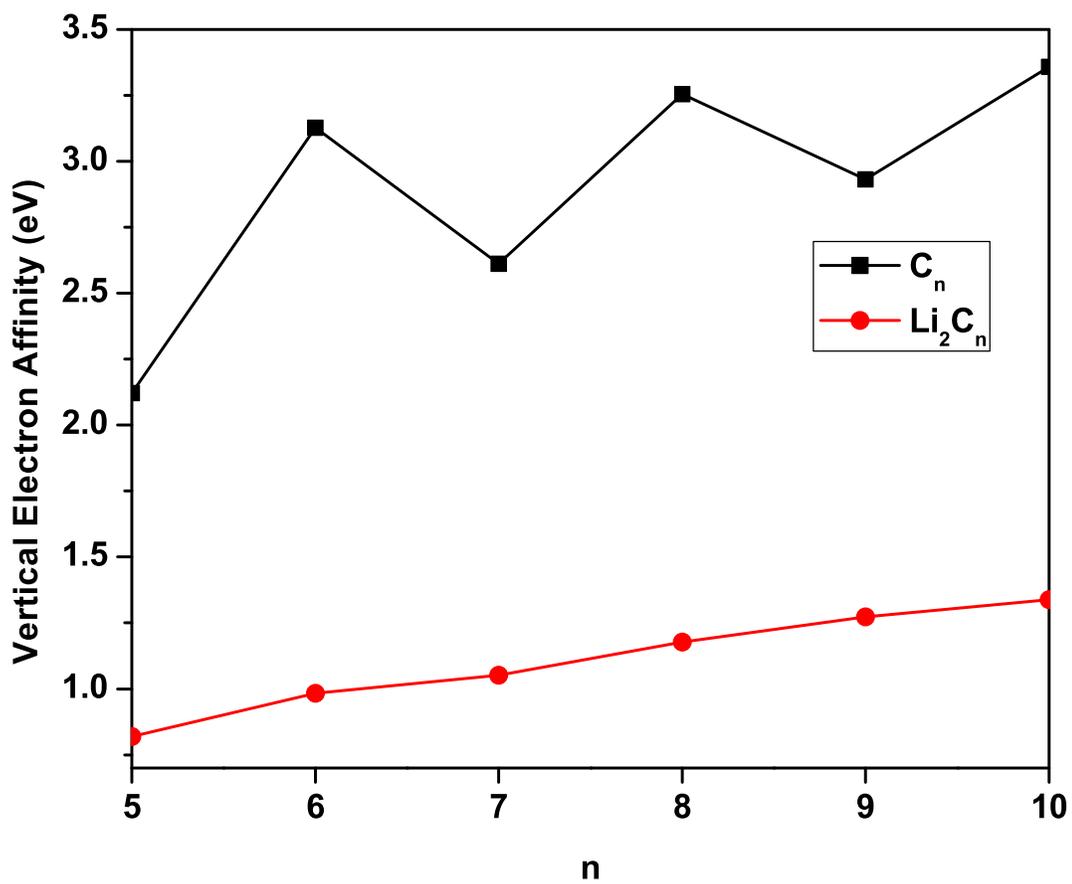} 
\caption{\label{fig:Figure_EAv} 
Vertical electron affinity for the lowest singlet state of 
C$_{n}$/Li$_{2}$C$_{n}$ as a function of the chain length, calculated using TAO-BLYP-D.} 
\end{figure}

\begin{figure} 
\includegraphics[scale=0.66]{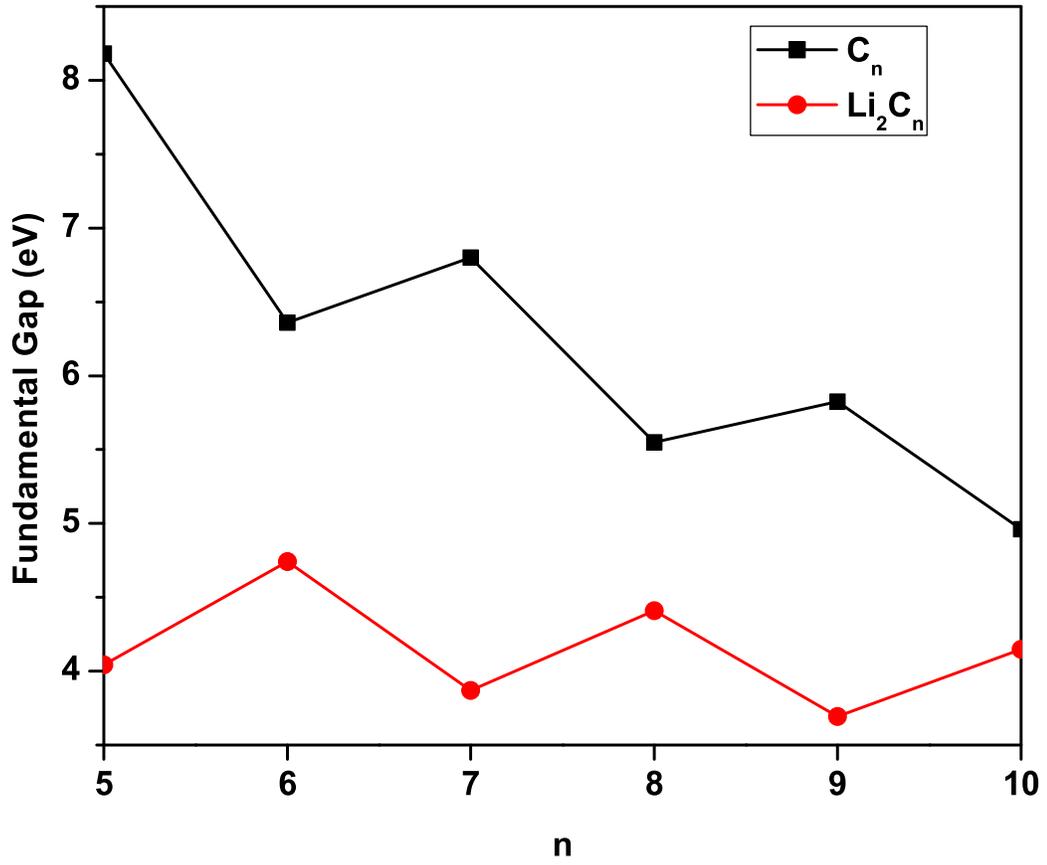} 
\caption{\label{fig:Figure_FG} 
Fundamental gap for the lowest singlet state of 
C$_{n}$/Li$_{2}$C$_{n}$ as a function of the chain length, calculated using TAO-BLYP-D.} 
\end{figure}

At the ground-state (i.e., the lowest singlet state) geometry of C$_{n}$/Li$_{2}$C$_{n}$ (with $N$ electrons), the vertical ionization potential ($\text{IP}_{v} = {E}_{N-1} - {E}_{N}$), 
vertical electron affinity ($\text{EA}_{v} = {E}_{N} - {E}_{N+1}$), and fundamental gap ($E_{g} = \text{IP}_{v} - \text{EA}_{v} = {E}_{N+1} + {E}_{N-1} - 2{E}_{N}$) are obtained with 
multiple energy-difference calculations, with ${E}_{N}$ being the total energy of the $N$-electron system. For each $n$, Li$_{2}$C$_{n}$ possesses the smaller 
$\text{IP}_{v}$ (see \Cref{fig:Figure_IPv}), $\text{EA}_{v}$ (see \Cref{fig:Figure_EAv}), and $E_{g}$ (see \Cref{fig:Figure_FG}) values than C$_{n}$. Note also that 
the $\text{IP}_{v}$, $\text{EA}_{v}$, and $E_{g}$ values of Li$_{2}$C$_{n}$ are less sensitive to the chain length than those of C$_{n}$. 

\begin{figure} 
\includegraphics[scale=0.66]{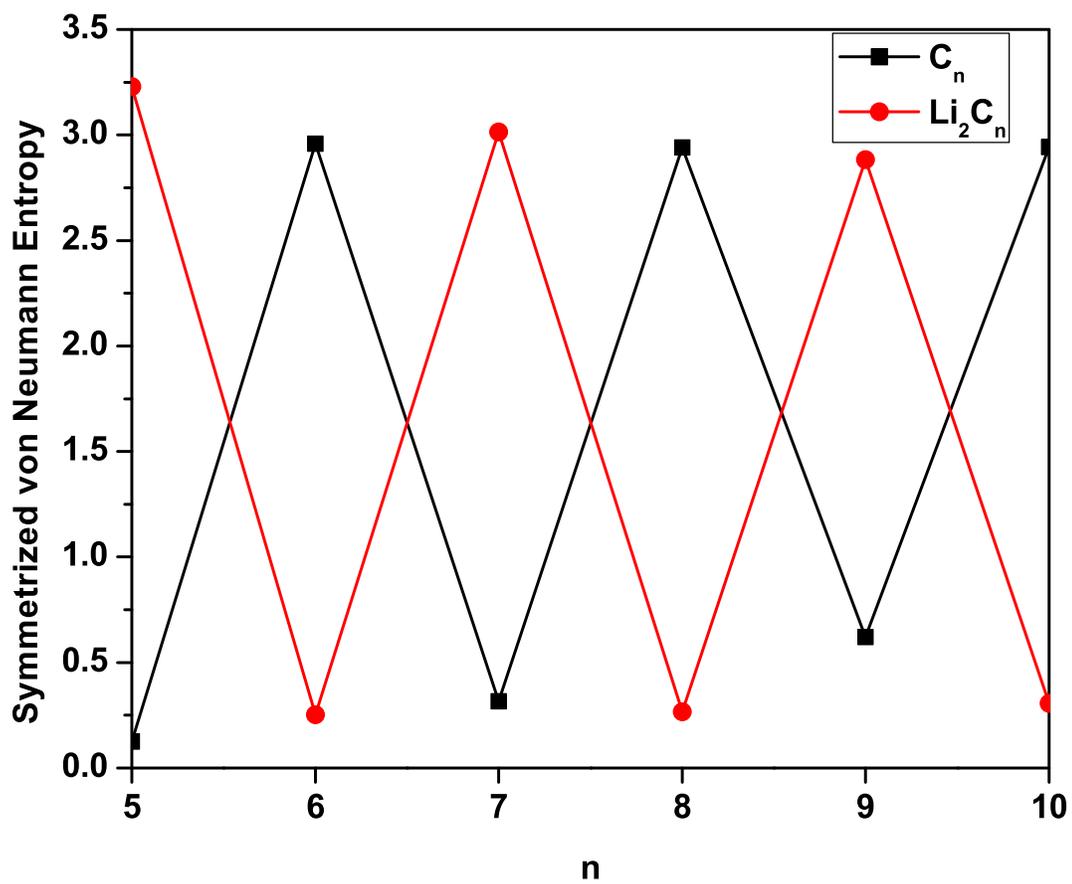} 
\caption{\label{fig:Figure_SvN} 
Symmetrized von Neumann entropy for the lowest singlet state of 
C$_{n}$/Li$_{2}$C$_{n}$ as a function of the chain length, calculated using TAO-BLYP-D.} 
\end{figure}

To examine the possible radical character of C$_{n}$/Li$_{2}$C$_{n}$, we calculate the symmetrized von Neumann entropy (e.g., see Eq.\ (9) of Ref.\ \cite{Rivero2013}) 
\begin{equation}\label{eq:svn} 
S_{\text{vN}} = -\frac{1}{2} \sum_{i=1}^{\infty} \bigg\lbrace f_{i}\ \text{ln}(f_{i}) + (1-f_{i})\ \text{ln}(1-f_{i}) \bigg\rbrace 
\end{equation} 
for the lowest singlet state of C$_{n}$/Li$_{2}$C$_{n}$ as a function of the chain length, using TAO-BLYP-D. Here, $f_{i}$ the occupation number of the $i^{\text{th}}$ orbital obtained with 
TAO-BLYP-D, which varies from 0 to 1, is approximately equal to the occupation number of the $i^{\text{th}}$ natural orbital \cite{ChaiTAO2012,ChaiTAO2014,ChaiTAO2017,noon}. 
For a system without strong static correlation ($\{f_{i}\}$ are close to either 0 or 1), $S_{\text{vN}}$ provides insignificant contributions, while for a system with strong static correlation 
($\{f_{i}\}$ are fractional for active orbitals, and are close to either 0 or 1 for others), $S_{\text{vN}}$ increases with the number of active orbitals. 
As shown in \Cref{fig:Figure_SvN}, the $S_{\text{vN}}$ values of 
C$_{n}$ with even-number carbon atoms and Li$_{2}$C$_{n}$ with odd-number carbon atoms are much larger than the $S_{\text{vN}}$ values of 
C$_{n}$ with odd-number carbon atoms and Li$_{2}$C$_{n}$ with even-number carbon atoms, respectively. 

\begin{figure} 
\includegraphics[scale=0.66]{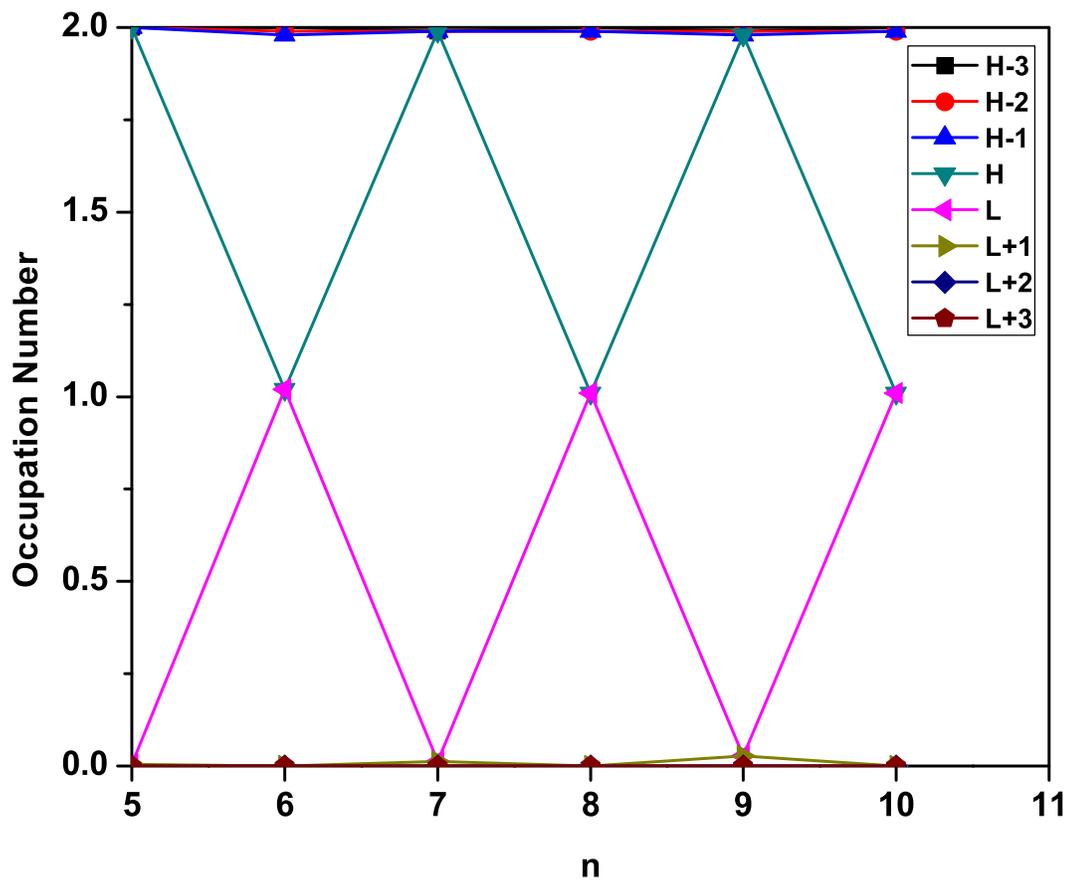} 
\caption{\label{fig:Figure_ON_Pure} 
Occupation numbers of the active orbitals (HOMO$-$3, HOMO$-$2, HOMO$-$1, HOMO, LUMO, LUMO+1, LUMO+2, and LUMO+3) 
for the lowest singlet state of C$_{n}$ as a function of the chain length, calculated using TAO-BLYP-D. 
For brevity, HOMO is denoted as H, LUMO is denoted as L, and so on.} 
\end{figure}

\begin{figure} 
\includegraphics[scale=0.66]{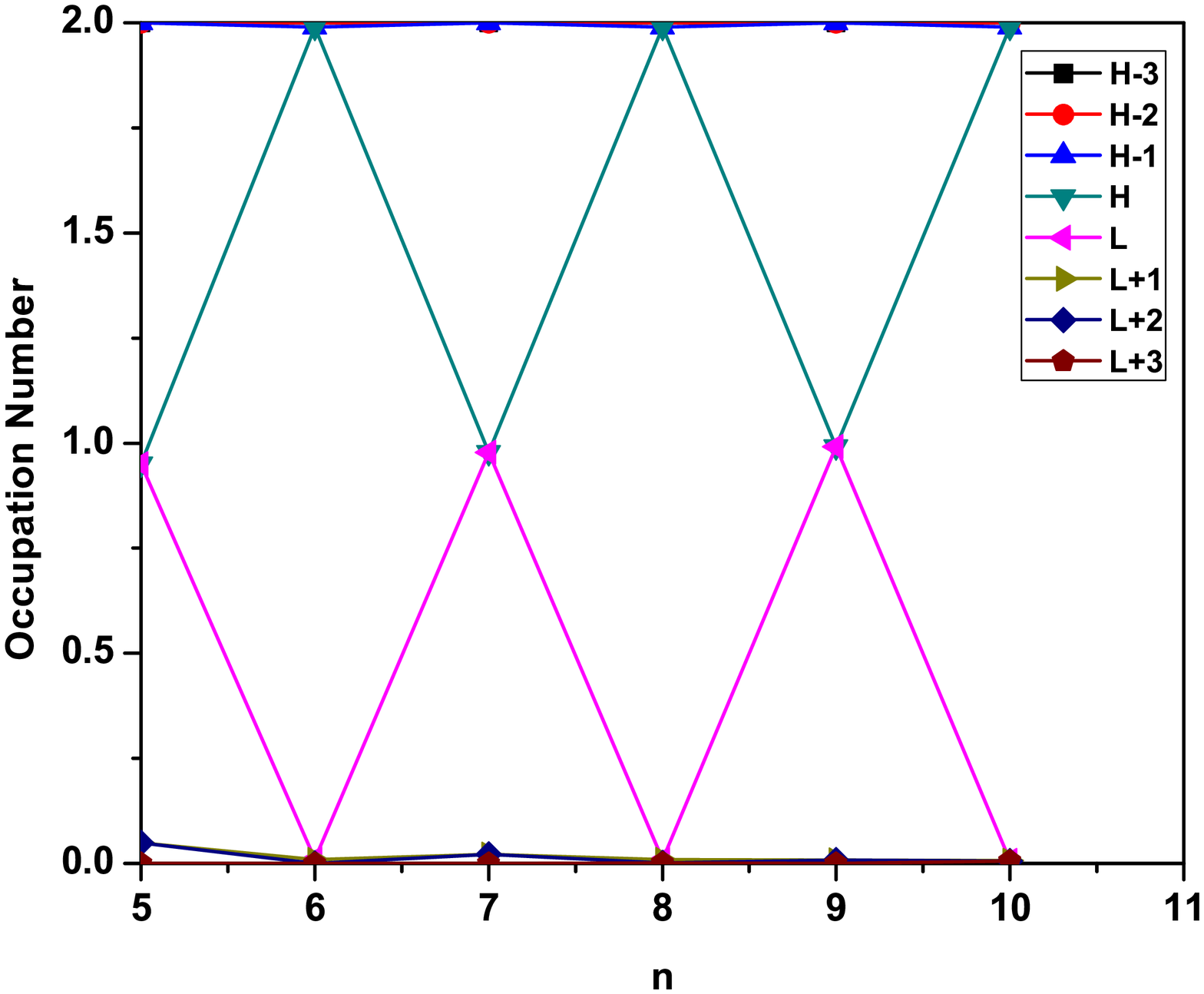} 
\caption{\label{fig:Figure_ON_Li2Cn} 
Same as \Cref{fig:Figure_ON_Pure}, but for Li$_{2}$C$_{n}$.} 
\end{figure}

To illustrate the causes of the odd-even oscillations in $S_{\text{vN}}$, we plot the occupation numbers of the active orbitals for the lowest singlet states of 
C$_{n}$ (see \Cref{fig:Figure_ON_Pure}) and Li$_{2}$C$_{n}$ (see \Cref{fig:Figure_ON_Li2Cn}), calculated using TAO-BLYP-D. Here, 
the highest occupied molecular orbital (HOMO) is the ${(N/2)}^{\text{th}}$ orbital, and the lowest unoccupied molecular orbital (LUMO) is the ${(N/2 + 1)}^{\text{th}}$ orbital, 
with $N$ being the number of electrons in C$_{n}$/Li$_{2}$C$_{n}$. For brevity, 
HOMO, HOMO$-$1, HOMO$-$2, and HOMO$-$3, are denoted as H, H$-$1, H$-$2, and H$-$3, respectively, while 
LUMO, LUMO+1, LUMO+2, and LUMO+3, are denoted as L, L+1, L+2, and L+3, respectively. As shown, 
C$_{n}$ with even-number carbon atoms and Li$_{2}$C$_{n}$ with odd-number carbon atoms possess more pronounced diradical character than 
C$_{n}$ with odd-number carbon atoms and Li$_{2}$C$_{n}$ with even-number carbon atoms, respectively. 

On the basis of several measures (e.g., the smaller ST gap, smaller $E_{g}$, larger $S_{\text{vN}}$, and more pronounced diradical character), 
C$_{n}$ with even-number carbon atoms and Li$_{2}$C$_{n}$ with odd-number carbon atoms should exhibit much stronger static correlation effects than 
C$_{n}$ with odd-number carbon atoms and Li$_{2}$C$_{n}$ with even-number carbon atoms (i.e., possessing single-reference character), respectively. 
Note that KS-DFT with conventional XC density functionals can be unreliable for the properties of systems with strong static correlation effects, and accurate multi-reference 
calculations are prohibitively expensive for large systems (e.g., the longer C$_{n}$ and Li$_{2}$C$_{n}$). In addition, due to the alteration of the reactivity of C$_{n}$ and 
Li$_{2}$C$_{n}$ with $n$, it is highly desirable to adopt an electronic structure method that can provide a balanced performance for both single- and multi-reference systems, 
well justifying the use of TAO-DFT in this study.

\subsection*{Hydrogen Storage Properties} 

As pure carbon materials bind H$_{2}$ molecules very weakly (i.e., mainly governed by vdW interactions), they are unlikely to be promising hydrogen storage materials 
at ambient conditions \cite{Bhatia2006}. Similarly, C$_{n}$ are not ideal for ambient storage applications, since the binding energies of H$_{2}$ molecules remain small. 
In addition, the number of H$_{2}$ molecules that can be adsorbed on C$_{n}$ is quite limited, due to the repulsive interaction between the adsorbed H$_{2}$ molecules 
at short distances \cite{Okamoto2001}. Consequently, the more the adsorbed H$_{2}$ molecules, the less the average H$_{2}$ binding energy on C$_{n}$. Therefore, 
C$_{n}$ cannot be high-capacity hydrogen storage materials at ambient conditions. 

\begin{figure} 
\includegraphics[scale=0.66]{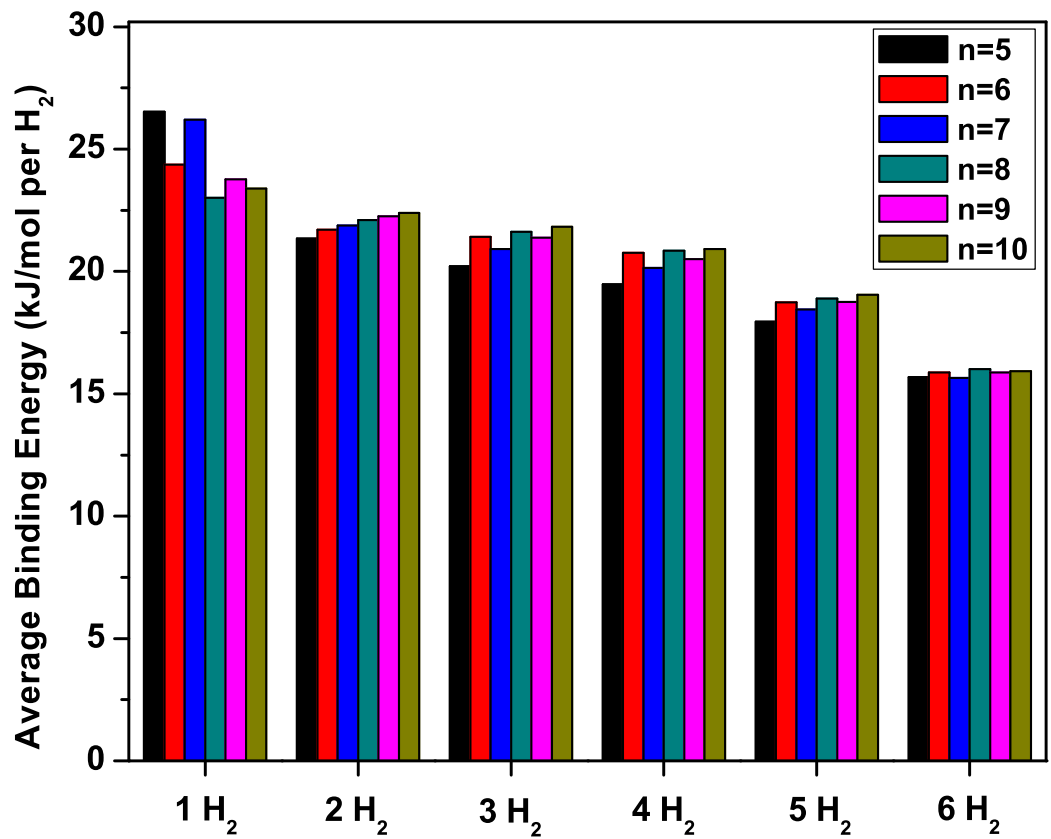} 
\caption{\label{fig:Figure_Avg_H2_Binding_Energy} 
Average H$_{2}$ binding energy on Li$_{2}$C$_{n}$ ($n$ = 5--10) as a function of the number of H$_{2}$ molecules adsorbed on each Li atom, calculated using TAO-BLYP-D.} 
\end{figure}

Here, we investigate the hydrogen storage properties of Li$_{2}$C$_{n}$ ($n$ = 5--10). As illustrated in \Cref{fig:Figure_Geometry}(b--h), at the ground-state geometry of 
Li$_{2}$C$_{n}$, $x$ H$_{2}$ molecules ($x$ = 1--6) are initially placed on various possible sites around each Li atom, and the structures are subsequently optimized 
to obtain the most stable geometry. All the H$_{2}$ molecules are found to be adsorbed molecularly to the Li atoms. The average H$_{2}$ binding energy, $E_{b}(\text{H}_{2})$, 
on Li$_{2}$C$_{n}$ is evaluated by 
\begin{equation}\label{eq:EBH2} 
E_{b}(\text{H}_{2}) = (E_{\text{Li}_{2}\text{C}_{n}} + 2x E_{\text{H}_{2}} - E_{\text{Li}_{2}\text{C}_{n}\text{-}2x\text{H}_{2}}) / (2x), 
\end{equation} 
where $E_{\text{H}_{2}}$ is the total energy of H$_{2}$, and $E_{\text{Li}_{2}\text{C}_{n}\text{-}2x\text{H}_{2}}$ is the total energy of Li$_{2}$C$_{n}$ with $x$ H$_{2}$ molecules 
adsorbed on each Li atom. Subsequently, $E_{b}(\text{H}_{2})$ is corrected for BSSE using a standard counterpoise correction \cite{Boys1970}. As shown 
in \Cref{fig:Figure_Avg_H2_Binding_Energy}, $E_{b}(\text{H}_{2})$ is in the range of 19 to 27 kJ/mol per H$_{2}$ for $x$ = 1--4, 
in the range of 18 to 19 kJ/mol per H$_{2}$ for $x$ = 5, and about 16 kJ/mol per H$_{2}$ for $x$ = 6, falling in (or close to) the ideal binding energy range. 

\begin{figure} 
\includegraphics[scale=0.66]{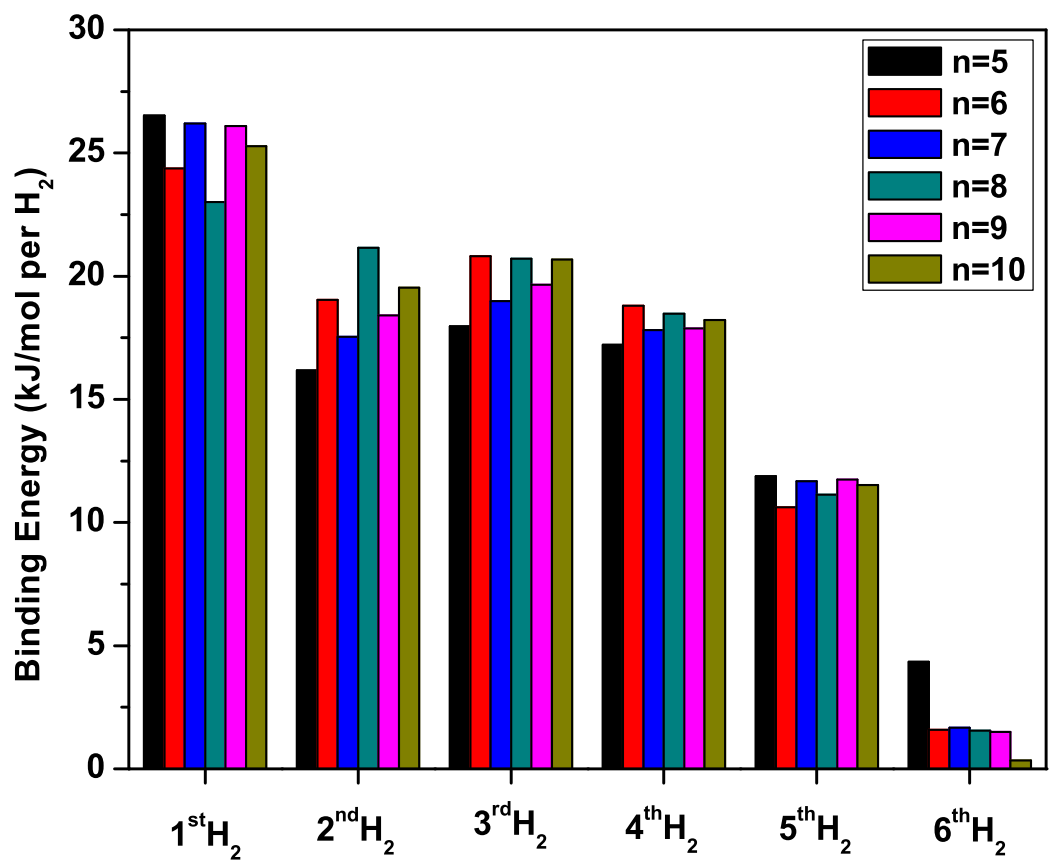} 
\caption{\label{fig:Figure_yth_H2_Binding_Energy} 
Binding energy of the $y^{\text{th}}$ H$_{2}$ molecule ($y$ = 1--6) on Li$_{2}$C$_{n}$ ($n$ = 5--10), calculated using TAO-BLYP-D.} 
\end{figure}

To assess if the binding energies of successive H$_{2}$ molecules are also in (or close to) the ideal binding energy range (i.e., not just the average H$_{2}$ binding energy), 
the binding energy of the $y^{\text{th}}$ H$_{2}$ molecule ($y$ = 1--6), $E_{b,y}(\text{H}_{2})$, on Li$_{2}$C$_{n}$ is evaluated by 
\begin{equation}\label{eq:EBH2add} 
E_{b,y}(\text{H}_{2}) = (E_{\text{Li}_{2}\text{C}_{n}\text{-}2(y-1)\text{H}_{2}} + 2 E_{\text{H}_{2}} - E_{\text{Li}_{2}\text{C}_{n}\text{-}2y\text{H}_{2}}) / 2. 
\end{equation} 
Similarly, $E_{b,y}(\text{H}_{2})$ is also corrected for BSSE using a standard counterpoise correction \cite{Boys1970}. As shown in \Cref{fig:Figure_yth_H2_Binding_Energy}, 
$E_{b,y}(\text{H}_{2})$ is in the range of 16 to 27 kJ/mol per H$_{2}$ for $y$ = 1--4, in the range of 11 to 12 kJ/mol per H$_{2}$ for $y$ = 5, and 
less than 5 kJ/mol per H$_{2}$ for $y$ = 6. Therefore, while the first four H$_{2}$ molecules can be adsorbed on Li$_{2}$C$_{n}$ in (or close to) the ideal binding energy range, 
the fifth and sixth H$_{2}$ molecules are only weakly adsorbed (i.e., not appropriate for ambient temperature storage). 

As Li$_{2}$C$_{n}$ ($n$ = 5--10) can bind up to 8 H$_{2}$ molecules (i.e., each Li atom can bind up to 4 H$_{2}$ molecules) with the average and successive H$_{2}$ binding 
energies in (or close to) the ideal binding energy range, the corresponding H$_{2}$ gravimetric storage capacity, $C_{g}$, is calculated using 
\begin{equation}\label{eq:Cg} 
C_{g} = \frac{8 M_{\text{H}_{2}}}{M_{\text{Li}_{2}\text{C}_{n}} + 8 M_{\text{H}_{2}}}. 
\end{equation} 
Here, $M_{\text{Li}_{2}\text{C}_{n}}$ is the mass of Li$_{2}$C$_{n}$, and $M_{\text{H}_{2}}$ is the mass of H$_{2}$. Note that $C_{g}$ (see Eq.\ (\ref{eq:Cg})) is 17.9 wt\% for $n$ = 5, 
15.8 wt\% for $n$ = 6, 14.1 wt\% for $n$ = 7, 12.8 wt\% for $n$ = 8, 11.7 wt\% for $n$ = 9, and 10.7 wt\% for $n$ = 10, satisfying the USDOE ultimate target of 7.5 wt\%. Based on the 
observed trends for Li$_{2}$C$_{n}$, the maximum number of H$_{2}$ molecules that can be adsorbed on each Li atom with the average and successive H$_{2}$ binding energies in 
(or close to) the ideal binding energy range should be 4, regardless of the chain length. Therefore, the $C_{g}$ value of Li$_{2}$C$_{n}$ should decrease as the chain length increases. 
Note, however, that the $C_{g}$ values obtained here may not be directly compared to the USDOE target value, which refers to the complete storage system (i.e., with the storage 
material, enclosing tank, insulation, piping, etc.) \cite{usdoe}. Nevertheless, since the $C_{g}$ values obtained here are much higher (especially for the shorter Li$_{2}$C$_{n}$) than 
the USDOE ultimate target, the complete storage systems based on Li$_{2}$C$_{n}$ could serve as high-capacity hydrogen storage materials for reversible hydrogen uptake and 
release at near-ambient conditions.

\section*{Conclusions} 

In conclusion, the search for ideal hydrogen storage materials have been extended to large systems with strong static correlation effects (i.e., those beyond the reach of traditional 
electronic structure methods), due to recent advances in TAO-DFT. In this work, we have studied the electronic properties (i.e., the Li binding energies, ST gaps, vertical ionization 
potentials, vertical electron affinities, fundamental gaps, symmetrized von Neumann entropy, and active orbital occupation numbers) and hydrogen storage properties (i.e., the average 
H$_{2}$ binding energies, successive H$_{2}$ binding energies, and H$_{2}$ gravimetric storage capacities) of Li$_{2}$C$_{n}$ ($n$ = 5--10) using TAO-DFT. As Li$_{2}$C$_{n}$ with 
odd-number carbon atoms have been shown to possess pronounced diradical character, KS-DFT with conventional XC density functionals can be unreliable for studying the properties 
of these systems. In addition, accurate multi-reference calculations are prohibitively expensive for the longer Li$_{2}$C$_{n}$ (especially for geometry optimization), and hence, the 
use of TAO-DFT in this study is well justified. On the basis of our results, Li$_{2}$C$_{n}$ can bind up to 8 H$_{2}$ molecules (i.e., each Li atom can bind up to 4 H$_{2}$ molecules) 
with the average and successive H$_{2}$ binding energies in (or close to) the ideal range of about 20 to 40 kJ/mol per H$_{2}$. Accordingly, the H$_{2}$ gravimetric storage capacities 
of Li$_{2}$C$_{n}$ are in the range of 10.7 to 17.9 wt\%, satisfying the USDOE ultimate target of 7.5 wt\%. Consequently, Li$_{2}$C$_{n}$ can be high-capacity hydrogen storage 
materials at near-ambient conditions. 

For the practical realization of hydrogen storage based on Li$_{2}$C$_{n}$, Li$_{2}$C$_{n}$ may be adopted as building blocks. For example, we may follow the proposal of 
Liu {\it et al.} \cite{Liu2011a}, and consider connecting Li-coated fullerenes with Li$_{2}$C$_{n}$, which could also serve as high-capacity hydrogen storage materials. A systematic 
study of the electronic and hydrogen storage properties of these systems is essential, and may be considered for future work. Since linear carbon chains \cite{Jin2009,Chuvilin2009} 
and Pt-terminated linear carbon chains \cite{Kano2014} have been successfully synthesized, the realization of hydrogen storage materials based on Li$_{2}$C$_{n}$ should be feasible, 
and is now open to experimentalists.

\section*{Acknowledgements} 
This work was supported by the Ministry of Science and Technology of Taiwan (Grant No.\ MOST104-2628-M-002-011-MY3), National Taiwan University (Grant No.\ NTU-CDP-105R7818), 
the Center for Quantum Science and Engineering at NTU (Subproject Nos.:\ NTU-ERP-105R891401 and NTU-ERP-105R891403), and the National Center for Theoretical Sciences of Taiwan. 
S.S. would like to thank Kerwin Hui and Chih-Ying Lin for useful discussions.

\section*{Author Contributions} 
S.S. and J.-D.C. designed the project. S.S. performed the calculations. S.S. and J.-D.C. contributed to the data analysis and writing of the paper.

\section*{Additional Information} 
{\bf Competing financial interests:} The authors declare no competing financial interests.

\newpage 

\newpage 

\newpage 

\newpage 

\newpage 

\newpage 

\newpage 

\newpage 

\newpage 

\newpage 

\newpage 

\end{document}